\documentstyle[a4wide,12pt,epsf]{article}
\begin{document}
\begin{flushright}
\baselineskip=12pt
{SUSX-TH-99-005}\\
{RHCPP99-08T}\\
{hep-ph/9906507}\\
{June 1999}
\end{flushright}

\begin{center}
{\LARGE \bf SPARTICLE SPECTRUM IN M-THEORY WITH FIVE-BRANE 
DOMINANCE \\}
\vglue 0.35cm
{D.BAILIN$^{\clubsuit}$ \footnote
{D.Bailin@sussex.ac.uk}, G. V. KRANIOTIS$^{\spadesuit}$ \footnote
 {G.Kraniotis@rhbnc.ac.uk} and A. LOVE$^{\spadesuit}$ \\}
	{$\clubsuit$ \it  Centre for Theoretical Physics, \\}
{\it University of Sussex,\\}
{\it Brighton BN1 9QJ, U.K. \\}
{$\spadesuit$ \it  Centre for Particle Physics , \\}
{\it Royal Holloway and Bedford New College, \\}
{\it  University of London,Egham, \\}
{\it Surrey TW20-0EX, U.K. \\}
\baselineskip=12pt

\vglue 0.25cm
ABSTRACT
\end{center}

{\rightskip=3pc
\leftskip=3pc
\noindent
\baselineskip=20pt
The phenomenological implications of the M-theory limit 
in which supersymmetry is broken by the F-terms of five-brane 
moduli is investigated.
In particular 
we calculate the supersymmetric spectrum subject to constraints of correct 
electroweak symmetry breaking. We find interesting differences especially 
in the squark sector compared with $M$-theory scenarios with standard 
embedding and weakly-coupled Calabi-Yau compactifications in the 
large $T$-limit. }

\vfill\eject
\setcounter{page}{1}
\pagestyle{plain}
\baselineskip=14pt
	
One of the most interesting developements in M-theory 
\cite{HORWIT,witten,ANTO,NANO,LIS,NILLES,DUDAS,STEVE,CHOI,LUKAS} 
model building 
is that new non-perturbative tools have been developed which 
allow the construction of realistic three generation models \cite{BURT}. 
In particular the inclusion of five-brane moduli $Z_n$, (which do not have 
a weakly coupled string theory counterpart) besides the 
metric moduli $T,S$ in the effective 
action leads to new types of $E_8 \times E_8$ symmetry breaking patterns  
as well as to novel gauge and K$\rm {\ddot a}$hler threshold 
corrections. As a result the soft-supersymmetry breaking terms 
differ substantially from the weakly coupled string.

Phenomenological implications of the effective action of M-theory 
with standard embedding of the spin connection into the gauge fields
have been investigated in \cite{BKL,KINE,CARLOS,CASA,LI}. 
Some phenomenological 
implications of non-standard embeddings in M-theory with and without 
five-branes have been 
studied in \cite{TATSUO,MUNOZ}. 
However, a full phenomenological analysis of the interesting case 
when the F-terms associated with the five-branes dominate  
those associated with metric moduli  
($F^{Z_n} \gg F^S,F^T$), including 
the constraint of radiative electroweak symmetry-breaking \cite{Tam:Rad} 
has not been performed. It is the purpose of this paper to 
carry out such an analysis.

The soft supersymmetry-breaking 
terms are determined by the following functions of the effective 
supergravity theory \cite{BURT,MUNOZ}:
\begin{eqnarray}
K&=&-{\rm ln}(S+\bar{S})-3{\rm ln}(T+\bar{T})+K_5+\frac{3}{T+\bar{T}}(
1+\frac{1}{3}e_O)H_{pq}C_O^p{\bar{C}^q_O}
, \nonumber \\
f_{O}&=&S+B_O T, \;\;f_{H}=S+B_H T, \nonumber \\
W_O&=&d_{pqr}C^p_O C^q_O C^r_O
\label{mfunc}
\end{eqnarray}
where $K$ is the K$\rm{\ddot{a}}$hler potential, $W$ the 
perturbative superpotential, and 
$f_{O}, f_{H}$ are the gauge kinetic functions for the observable 
and hidden sector gauge groups respectively.
$K_5$ is the K$\rm{\ddot a}$hler potential for the five-brane 
moduli $Z_n$ and $H_{pq}$ is some $T$-independent metric. 
Also
\begin{equation}
e_O=b_O \frac{T+\bar{T}}{S+\bar{S}},\;\;
e_H=b_H\frac{T+\bar{T}}{S+\bar{S}}
\end{equation}
The various instanton numbers are given by the foloowing expressions
\begin{eqnarray}
b_O &=&\beta_O+\sum_{n=1}^{N}(1-z_n)^2 \beta_n \nonumber \\
B_O &=&\beta_O+\sum_{n=1}^{N}(1-Z_n)^2 \beta_n \nonumber \\
B_H &=&\beta_H+\sum_{n=1}^{N}(Z_n)^2 \beta_n \nonumber \\
b_H &=&\beta_H+\sum_{n=1}^{N}z_n^2 \beta_n
\end{eqnarray}
Since a Calabi-Yau manifold is compact, the net magnetic charge 
due to orbifold planes and 5-branes is zero. Consequently the 
following cohomology condition is satisfied
\begin{equation}
\beta_O+\sum_{n=1}^{N}\beta_n+\beta_H=0
\end{equation}
$S,T$ are the dilaton and 
Calabi-Yau moduli fields and $C^{p}$ charged matter fields.
The five-brane moduli are denoted by $Z_n$, and
${\rm{Re}} Z_n \equiv z_n=\frac{x_n}{\pi \rho} \in (0,1)$.
The superpotential
and the gauge kinetic functions are exact up to non-perturbative effects.

Given eqs(\ref{mfunc}) one can determine 
\cite{BURT,MUNOZ} the soft supersymmetry breaking terms
for the 
observable sector 
gaugino masses $M_{1/2}$, scalar masses $m_0$ and trilinear scalar 
masses $A$
as functions of the auxiliary fields $F^S$,$F^T$, $F^{n}$ of the moduli 
$S,T$ fields and five-brane moduli $Z_n$ respectively.
\begin{eqnarray}
M_{1/2}&=&
\frac{1}{(S+\bar{S})(1+\frac{B_O T+\bar{B}_O \bar{T}}{S+\bar{S}})}
(F^S+F^T B_O+T F^n \partial_n B_O) \nonumber \\
m_0^2&=&V_0+m_{3/2}^2-\frac{1}{(3+e_O)^2}\Bigr[
e_O (6+e_O) \frac{|F^S|^2}{(S+\bar{S})^2} \nonumber \\
&+& 3(3+2e_O)\frac{|F^T|^2}{(T+\bar{T})^2}-
\frac{6 e_O}{(S+\bar{S})(T+\bar{T})} Re F^S {\bar{F}^{\bar{T}}} \nonumber \\
&+& \Bigl(\frac{e_O}{b_O}(3+e_O)\partial_n \partial_{\bar{m}} 
b_O-\frac{e_O^2}{b_O^2}\partial_n b_O 
\partial_{\bar{m}} b_O\Bigl) F^n \bar{F}^{\bar{m}} \nonumber \\
&-&\frac{6 e_O}{b_O}\frac{\partial_{\bar{n}} b_O}{S+\bar{S}}
Re F^S \bar{F}^{\bar{n}}+\frac{6 e_O}{b_O}
\frac{\partial_{\bar{n}} b_O}{T+\bar{T}} Re F^T \bar{F}^{\bar{n}}
\Bigr], \nonumber \\
A&=&-\frac{1}{3+e_O}\Bigl\{\frac{F^S (3-2 e_O)}{S+\bar{S}}+
\frac{3 e_O F^T}{T+\bar{T}} \nonumber \\
&+& F^n \Bigl(3\frac{e_O}{b_O}\partial_n b_O-(3+e_O)\partial_n K_5 \Bigr)
\Bigr\}
\label{GAUG}
\end{eqnarray}
where $\partial_n \equiv \frac{\partial}{\partial Z_n}$.
The  bilinear $B$-parameter  associated with non-perturbatively generated 
$\mu$ term in the superpotential 
is given by \cite{MUNOZ}:
\begin{eqnarray}
B_{\mu}&=& \frac{F^S (e_O-3)}{(3+e_O)(S+\bar{S})}-
\frac{3 (e_O+1) F^T}{(T+\bar{T})(3+e_O)} \nonumber \\
&+&\frac{1}{3+e_O}\Bigl[(3+e_O)F^n\partial _n K_5-2F^n 
\frac{e_O}{b_O}\partial_n b_O\Bigr]-m_{3/2} \nonumber \\
\label{beta}
\end{eqnarray}
From now on we assume that only one five-brane contributes to 
supersymmetry-breaking \footnote{We assume very small 
$CP$-violating phases in the soft terms.}.
Then the  auxiliary fields are given by \cite{MUNOZ,IBA:Spain}
\begin{eqnarray}
F^1&=&\sqrt{3} m_{3/2} C (\partial_1 \partial_{\bar{1}} K_5)^{-1/2} 
\sin\theta_1 \nonumber \\
F^S &=& \sqrt{3} m_{3/2} C (S+\bar{S}) \sin\theta \cos\theta_1 \nonumber \\
F^T &=& m_{3/2} C (T+\bar{T}) \cos\theta \cos\theta_1 
\end{eqnarray}
The goldstino angles are denoted by $\theta,\theta_1$, $m_{3/2}$ is 
the gravitino mass and $C^2=1+\frac{V_0}{3 m_{3/2}^2}$ 
with $V_0$ the tree level vacuum 
energy density. The five-brane dominated supersymmetry-breaking 
scenario corresponds to $\theta_1=\frac{\pi}{2}$, i.e 
$F^T,F^S=0$, and we take    
the 
five brane which contributes to supersymmetry breaking to be located 
at $z_1=1/2$ in the orbifold interval. We also set $C=1$ in the 
above expressions assuming zero cosmological constant.

We now consider the supersymmetric particle spectrum for 
a single five-brane present.
Our parameters are, $e_O$, $\partial_1 \partial_{\bar{1}} K_5$,
$\partial_1 K_5$, 
$m_{3/2}$, $sign\; \mu$ (which is not determined by the radiative 
electroweak symmetry breaking constraint), where 
$\mu$ is the Higgs mixing parameter in the 
low energy superpotential. 
The ratio of the two Higgs vacuum 
expectation values $\tan\beta=\frac{<H_2^0>}{<H_1^0>}$ is also 
a free parameter if we leave  
$B$ be  determined 
by  the minimization of the one-loop Higgs effective potential. 
If $B$ instead is given by
(\ref{beta}), one determines the value of $\tan\beta$. 
For this purpose  we take $\mu$ independent of $T$ and $S$
because of our lack of knowledge of 
$\mu$ in $M$-theory.
We treat $e_O$ as a free parameter as the problem of 
stabilizing the dilaton and other moduli has not yet been solved, 
although there has been an interesting work in this area \cite{KOREA}.

The instanton numbers are model dependent. In this paper 
we  choose to work with the interesting example \cite{MUNOZ} with 
$\beta_O=-2$ and $\beta_1=1$ which implies $b_O=-7/4$. 
This implies that $b_H=5/4$ and allow us to study the region of 
parameter space with
$-1<e_O \leq 0$, which is not accessible in strongly coupled M-theory 
scenarios with standard embedding.
We also choose $\partial_1 K_5=\partial_1 \partial_{\bar{1}}K_5=1$.
We have investigated deviations from these values in order to 
examine the robustness of our results. We comment on that below.

We  use the following experimental bounds from unsuccessful 
searches at LEP and Tevatron for supersymmetric particles 
\cite{EUC}. 
We require the lightest 
chargino $M_{\chi^+_1}\geq 95$ GeV, 
and the lightest Higgs, $m_{h_0} >79.6$ GeV. A lower 
limit on the mass of the lightest stop  $m_{\tilde{t}_2} >86$ GeV, 
from $\tilde{t}_2 \rightarrow c 
\chi_1^0$ decay  in  D$0$ is imposed. The stau 
mass eigenstate ($\tilde{\tau}$) should be heavier 
than 53 GeV from LEP2 results.

The soft masses start running from 
a mass $R_{11}^{-1}\sim 7.5 \times 10^{15}$ GeV with 
$R_{11}$ the extra $M$-theory dimension. This is perhaps the most 
natural choice, although values as low as $10^{13}$ GeV are possible 
and have been advocated by some authors \cite{NANO}. 
However, the analysis of \cite{NILLES} disfavours such 
scenarios. For the most part of our analysis we shall therefore consider the 
former value of $R_{11}$, but we shall also comment
on the consequences of the latter.
Then using (\ref{GAUG}),(\ref{beta}) as boundary 
conditions for the soft terms,
one evolves the renormalization 
group equations down to the weak scale and determines 
the sparticle 
spectrum compatible with the constraints of correct electroweak symmetry 
breaking and the above experimental 
constraints on the sparticle 
spectrum.  

Electroweak symmetry breaking is characterized by the extrema equations 
\begin{eqnarray}
\frac{1}{2}M_Z^2&=&\frac{\bar{m}^2_{H_1}-\bar{m}^2_{H_2}\tan^2 \beta}
{\tan^2 \beta -1}-\mu^2 \nonumber \\
-B\mu&=&\frac{1}{2}(\bar{m}^2_{H_1}+\bar{m}^{2}_{H_2}+2\mu^2)\sin 2\beta
\end{eqnarray}
where 
\begin{equation}
\bar{m}^2_{H_1,H_2}\equiv m^2_{H_1,H_2}+\frac{\partial \Delta V}{
\partial {v^2_{1,2}}}
\end{equation}
and $\Delta V=(64 \pi^2)^{-1} {\rm {STr}} M^4[ln (M^2/Q^2)-\frac{3}{2}]$ 
is the 
one loop contribution to the Higgs effective potential. We include
contributions only from the third generation of particles and sparticles.

Since $\mu^2 \gg M_Z^2$ for most of the allowed
region of the parameter space \cite{NATH}, the following 
approximate relationships 
hold at the 
electroweak scale for the 
masses of neutralinos and charginos, which of 
course depend on the details of 
electroweak symmetry breaking. 
\begin{eqnarray}
m_{\chi_1^{\pm}}\sim m_{\chi_2^0}\sim 2 m_{\chi_1^0} \nonumber \\
m_{\chi_{3,4}^{0}}\sim m_{\chi_2^{\pm}}\sim |\mu|
\label{neutralinos}
\end{eqnarray}
In (\ref{neutralinos}) $m_{\chi_{1,2}^{\pm}}$ are the chargino mass 
eigenstates and $m_{\chi_{i}^{0}},i=1\ldots 4$ are the four neutralino mass 
eigenstates with $i=1$ 
denoting the lightest 
neutralino. The former arise after diagonalization of the 
mass matrix. 
\begin{equation}
M_{ch}=\left(\begin{array}{cc}\\
M_2 & \sqrt{2} m_W \sin\beta \\
m_W \cos\beta & -\mu 
\end{array}\right)
\end{equation}
The stau mass matrix is given by the expression 

\begin{equation}
{\cal M}_{\tau}^2=\left(\begin{array}{cc} \\
{\cal M}_{11}^2 & m_{\tilde{\tau}}(A_{\tau}+\mu \tan\beta) \\
m_{\tilde{\tau}}(A_{\tau}+\mu \tan\beta) & {\cal M}_{22}^2
\end{array}\right)
\end{equation}
where ${\cal M}_{11}=m^2_L+
m^2_{\tau}-
\frac{1}{2}(2M_W^2-M_Z^2)\cos 2 \beta$ and ${\cal M}_{22}=
m^2_E+m^2_{\tau}+(M_W^2-M_Z^2)\cos2\beta$.
where $m^2_L, \; m^2_E$ refer to scalar soft masses for lepton doublet, 
singlet respectively.


As has been noted in \cite{MUNOZ}, in the case when only the 
five-branes contribute to supersymmetry-breaking the ratio of 
scalar masses to gaugino mass, $m_0/|M_{1/2}|>1$ for $e_O>-0.65$.
This is quite interesting since scalar masses larger than gaugino masses 
are not easy to obtain in the weakly-coupled heterotic string or 
M-theory compactification with standard embedding. In the case of non-standard 
embeddings without five-branes there is small region 
of the parameter space where is possible 
to have  $m_0/|M_{1/2}|>1$. 

In fig.1 we present the sparticle spectrum plotted against 
$\tan\beta$ for $e_O=-0.6,\; m_{3/2}=165$ GeV and $\mu<0$, 
$\partial_1 K_5=\partial_1 \partial_{\bar{1}} K_5=1$. For this 
choice of $e_O$, $m_0/|M_{1/2}|\sim 1.2$ at the 
unification scale. The scalar masses are bigger 
than the   gaugino masses, in contrast to weakly coupled string 
scenarios emerging from 
the large $T$-limit of Calabi-Yau 
compactifications where gaugino masses are generically 
bigger than scalar masses. One can see that the lightest 
supersymmetric particle is the lightest neutralino.
It is a linear combination of the superpartners of the 
photon, $Z^0$ and neutral-Higgs bosons,
\begin{equation}
\chi_1^0=c_1 \tilde{B}+c_2\tilde{W}^3+c_3\tilde{H}_1^0+c_4\tilde{H}_2^0
\end{equation}
The neutralino $4\times 4$ mass matrix can be written as
$$\left(\begin{array}{cccc} \\
M_1 & 0 & -M_Z A_{11} & M_Z A_{21} \\
0  & M_2 & M_Z A_{12} &-M_Z A_{22} \\
-M_Z A_{11} & M_Z A_{12} & 0 & \mu \\
M_Z A_{21} & -M_Z A_{22} & \mu & 0
\end{array}\right)$$ 
with 
$$\left(\begin{array}{cc} \\
 A_{11} & A_{12} \\
A_{21} & A_{22}\end{array}\right)= 
\left(\begin{array}{cc} \\
\sin\theta_{W} \cos\beta & \cos \theta_W \cos\beta \\
\sin\theta_W \sin\beta & \cos\theta_W \sin\beta 
\end{array}\right)$$

The next to lightest supersymmetric particle (NLSP) in this scenario  is 
the lightest chargino. In this figure we plot chargino masses 
heavier than the experimental bound from LEP, $m_{\chi_1^+}\geq 95$GeV.
In fact this experimental bound imposes $m_{3/2}>160$ GeV.
The lightest Higgs eigenstate ($h$) satisfies  $m_h> 85$ GeV 
for $\tan\beta=2.5$ and $m_h\geq 109$ GeV for 
$\tan\beta=10$. Also  $m_{\tilde{t}_2}\geq 183$ GeV for 
all values of $\tan\beta$ and the lightest 
stau satisfies  $m_{\tilde{\tau}_2}\geq 127$ GeV.
The lightest stau mass eigenstate is the next to the NLSP (NNLSP)  
mass eigenstate. This is to be contrasted  with the 
extreme M-theory limit with standard embedding for 
a goldstino angle chosen to get 
vanishing scalar masses at the unification 
scale. In the latter scenario 
the lightest stau is the NLSP and for values 
of $\tan\beta \geq 12$ becomes the LSP. 
It differs from the $M$-theory scenario with goldstino 
angle $\frac{\pi}{8}$ and $\tan\beta \geq 10$ since in this case again 
the lightest stau is the NLSP. 
It also differs from these M-theory scenarios in the squark sector.
In particular, the lightest stop is always much lighter 
than the CP odd Higgs mass eigenstate$A^0$. This is not the generic 
case with $M$-theory scenarios with standard embeddings 
\cite{BKL} and 
weakly coupled CY compactifications in the large $T$-limit \cite{IBA:Spain}.  
Similarly the lightest sbottom $\tilde{b}_2$ (sb2 in the figure) 
is almost degenerate with the CP odd Higgs mass (is lighter 
for small $\tan\beta$ and heavier for large $\tan\beta$ )in contrast to 
M-theory scenarios with standard embedding and various goldstino 
angles. In the latter case the lightest sbottom is much heavier than the 
CP-odd Higgs mass \cite{BKL,KINE} in most region of the parameter space. 
We also 
observe from figure 1 that the 
lightest stop is lighter than the lightest sbottom mass eigenstate. 
For $\tan\beta\leq 2 $ and for this particular gravitino mass 
the lightest Higgs becomes lighter than the current experimental 
bound of $79.6$ GeV. 
In figure 3 we present the sparticle spectrum plotted against 
$\tan\beta$ 
for $\mu>0$.  Throught the region of parameter space the 
lightest neutralino is the LSP and the lightest chargino the NLSP.
The lightest stau is the NNLSP. 

In figure 2 we present the sparticle spectrum with respect to the 
parameter $e_O$ for $\tan\beta=2.5$, and $m_{3/2}=250$ GeV. We note  
the following very interesting features. As the 
parameter $e_O$ increases 
through $-0.65$
the scalar masses become heavier than the gaugino masses. The imprint 
in the low energy spectra is that  
the lightest chargino is the 
NLSP, and for $e_O>-0.5$ the lightest stop becomes the NNLSP. 
In the limit $e_O \rightarrow 0$ the scalar masses 
are much heavier than gaugino masses, $m_0\gg M_{1/2}$. 
For smaller (more negative) values of $e_O$  the 
lightest stau becomes lighter  as the gaugino masses become 
heavier  and eventually for $e_O\leq -0.7$ becomes the NLSP.
For instance for $e_O=-0.7, |M_{1/2}|/m_O\sim 1.32$ while for 
$e_O=-0.8, |M_{1/2}|/m_O \sim 2.33$.
In the region  $-1<e_O<-0.7$, the  spectrum 
resembles strongly  coupled $M$-theory scenarios  
without five-branes for some values of the 
goldstino angle, e.g.  $\theta=\frac{7\pi}{20}$. In contrast for 
this particular $\tan\beta$
in $M$-theory scenarios without five branes and for 
$\theta=\frac{\pi}{8}$ the lightest chargino is the NLSP.

When the $B_{\mu}$ term is given by (\ref{beta}) correct electroweak 
symmetry breaking occurs for $18\leq \tan\beta \leq 20.5$ in the region 
$-0.7\leq e_O \leq -0.2$.
In fig.4 we present the supersymmetric particle spectrum with 
respect to $e_O$ for $\tan\beta=20$ and $m_{3/2}=230$ GeV. The same 
qualitative features hold as in the case of low $\tan\beta$ discussed 
earlier. The neutralino is the LSP in most of the parameter space 
($-0.85<e_O<0$). For $e_O>-0.65$ the lightest chargino is the NLSP 
while for $-0.85<e_O<-0.7$ the lightest stau is the NLSP. For 
$e_O\geq -0.5$ the lightest stop is the NNLSP.

We also investigated the supersymmetric particle spectrum for 
$\partial_1 \partial_{\bar{1}} K_5=2,0.5$. The same qualititative 
features discussed earlier hold also for this choice of the 
five-brane Kahler potential metric.

In conclusion we have calculated the supersymmetric particle spectrum in 
the five-brane dominated limit of M-theory subject to the 
constraint of correct electroweak symmetry breaking and experimental 
bounds from accelerator physics. 
For most of the parameter space the lightest 
neutralino is the LSP. Depending on the value of 
$e_O$ and $\tan\beta$ the NLSP is either the lightest chargino or the 
the lightest stau. In the region of parameter space for which 
$m_0/|M_{1/2}|>1$ the lightest chargino is the NLSP. For $\tan\beta=2.5$, 
$m_{3/2}=250$ GeV, and for $e_O>-0.5$ the lightest stop becomes the 
NNLSP. For $e_O<-0.5$ the lightest stau is the NNLSP and eventually 
as  $e_O$ decreases becomes the NLSP.   
In the region of  parameter space in which $m_0>|M_{1/2}|$ 
the five-brane dominated scenario leads to qualitative differences 
from $M$-theory scenarios with standard embeddings and Calabi-Yau 
compactifications in the large $T$-limit, especially in the 
squark sector.
The resulting particle spectra should be subject of experimental 
investigation.
They also motivate the study of how well the dark matter constraints 
are satisfied. We expect that for large values of the $e_O$ close to $0$ 
the dark matter constraints can become quite important in setting 
upper limits on the gluino mass. Also in this region the higgsino 
component of the lightest neutralino increases and this can be 
quite interesting for direct dark matter detection. These matters 
will be a subject of a future publication \cite{MAS}.

\section*{Acknowledgements}
This research is supported in part by PPARC.

\newpage

\begin{figure}
\epsfxsize=6.5in
\epsfysize=8.5in
\epsffile{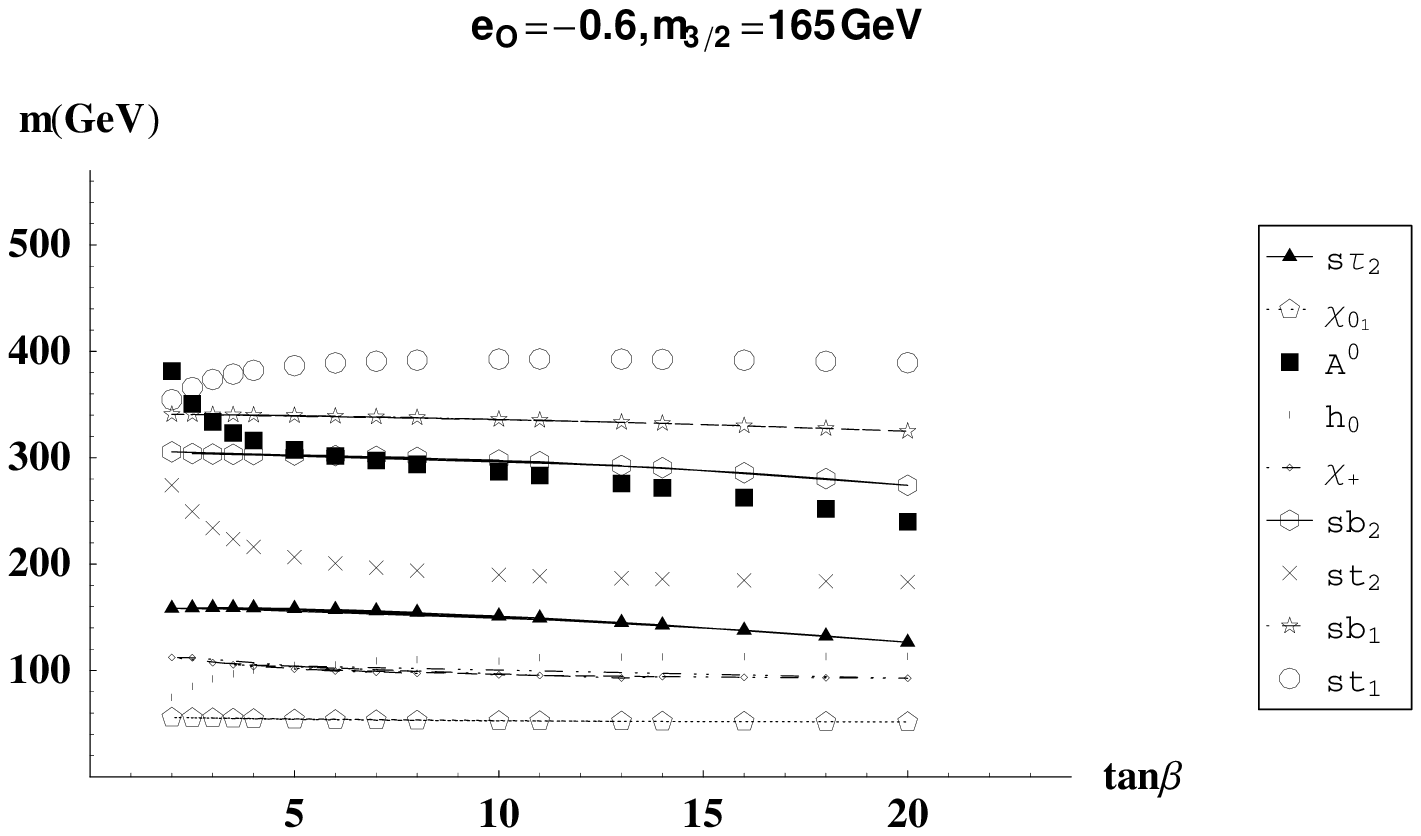}
\caption{Sparticle spectrum vs $\tan\beta$ for $e_O=-0.6,\partial_1 K_5=
\partial_1 \partial_{\bar{1}} K_5=1,m_{3/2}=165GeV \mu<0$.}
\end{figure}

\begin{figure}
\epsfxsize=6in
\epsfysize=8.3in
\epsffile{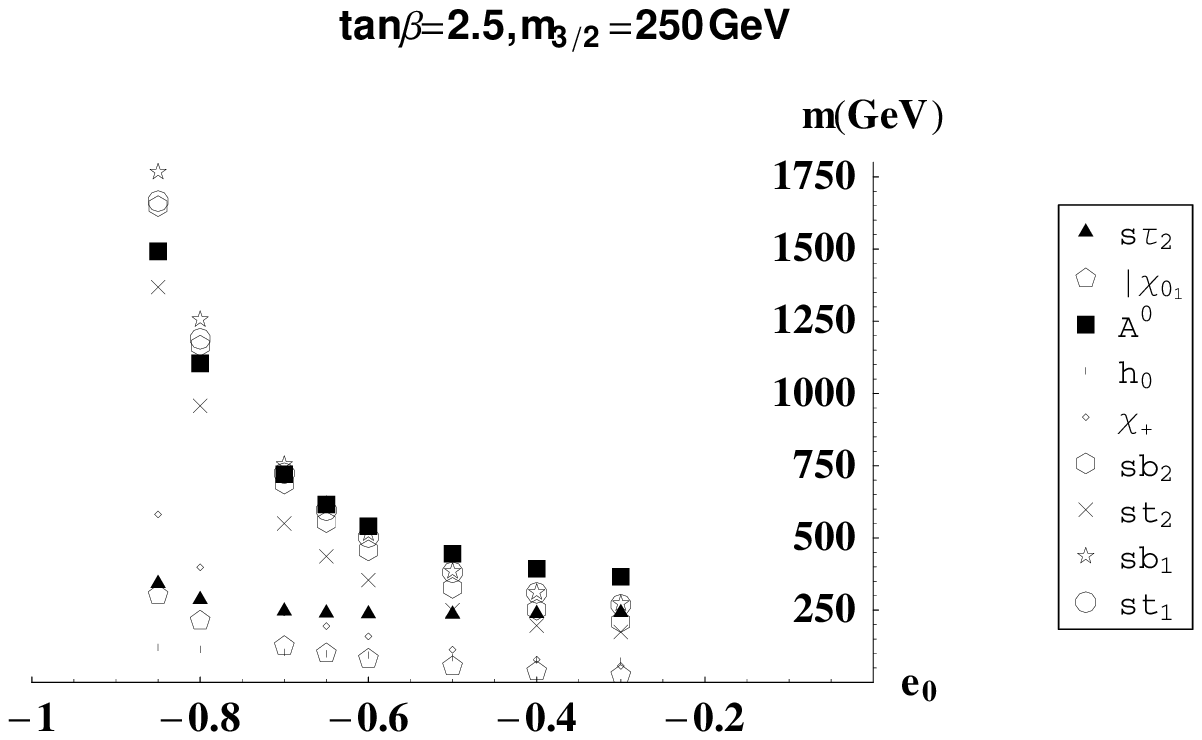}
\caption{Sparticle spectrum vs $e_O$ for fixed $\tan\beta=2.5,
m_{3/2}=250GeV$.}
\end{figure}

\newpage
\begin{figure}
\epsfxsize=6.8in
\epsfysize=9.2in
\epsffile{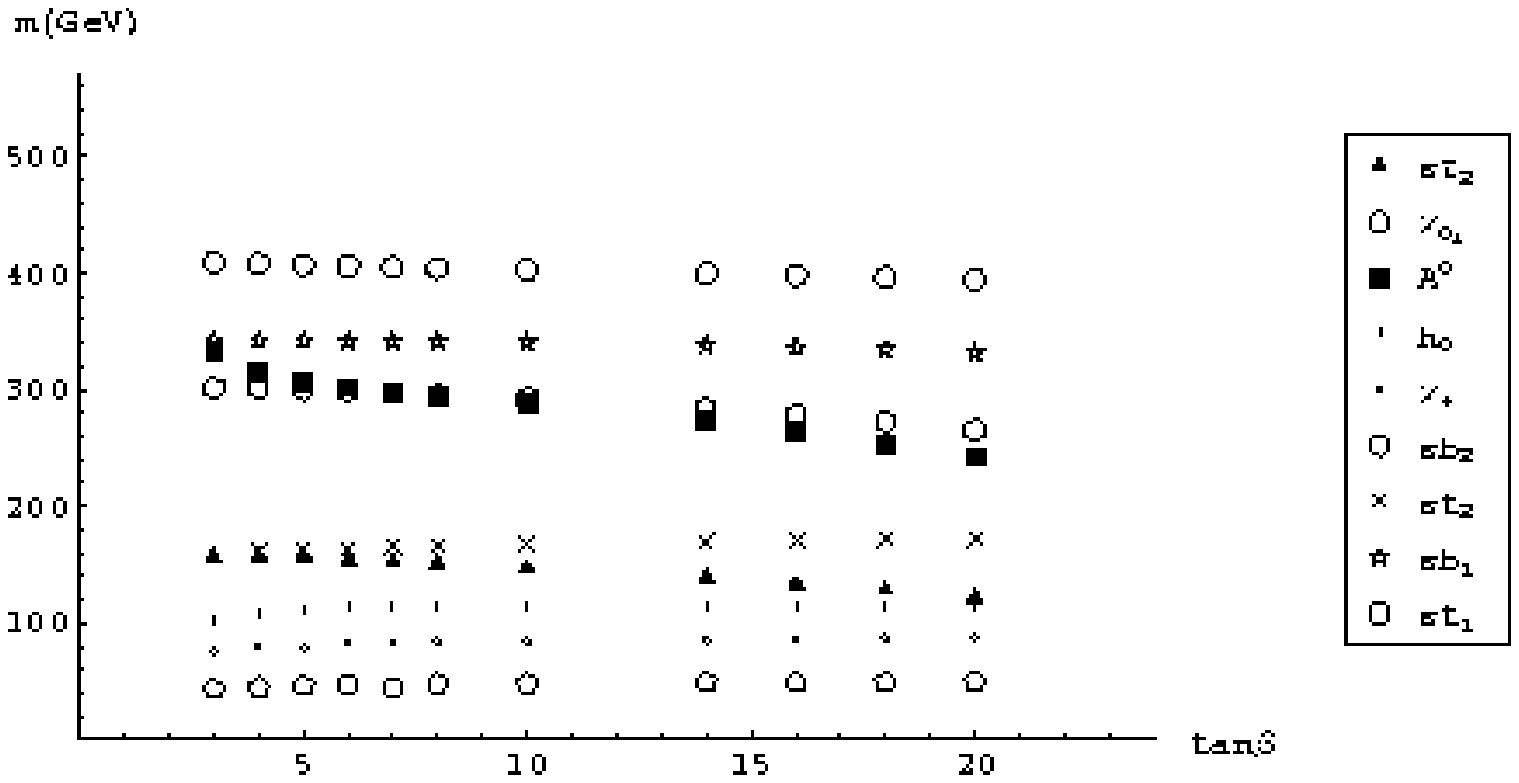}
\caption{Sparticle spectrum vs $\tan\beta$ for $e_O=-0.6,\partial_1 K_5=
\partial_1 \partial_{\bar{1}} K_5=1,m_{3/2}=165GeV, \mu>0$.}
\end{figure}

\newpage
\begin{figure}
\epsfxsize=6in
\epsfysize=8.5in
\epsffile{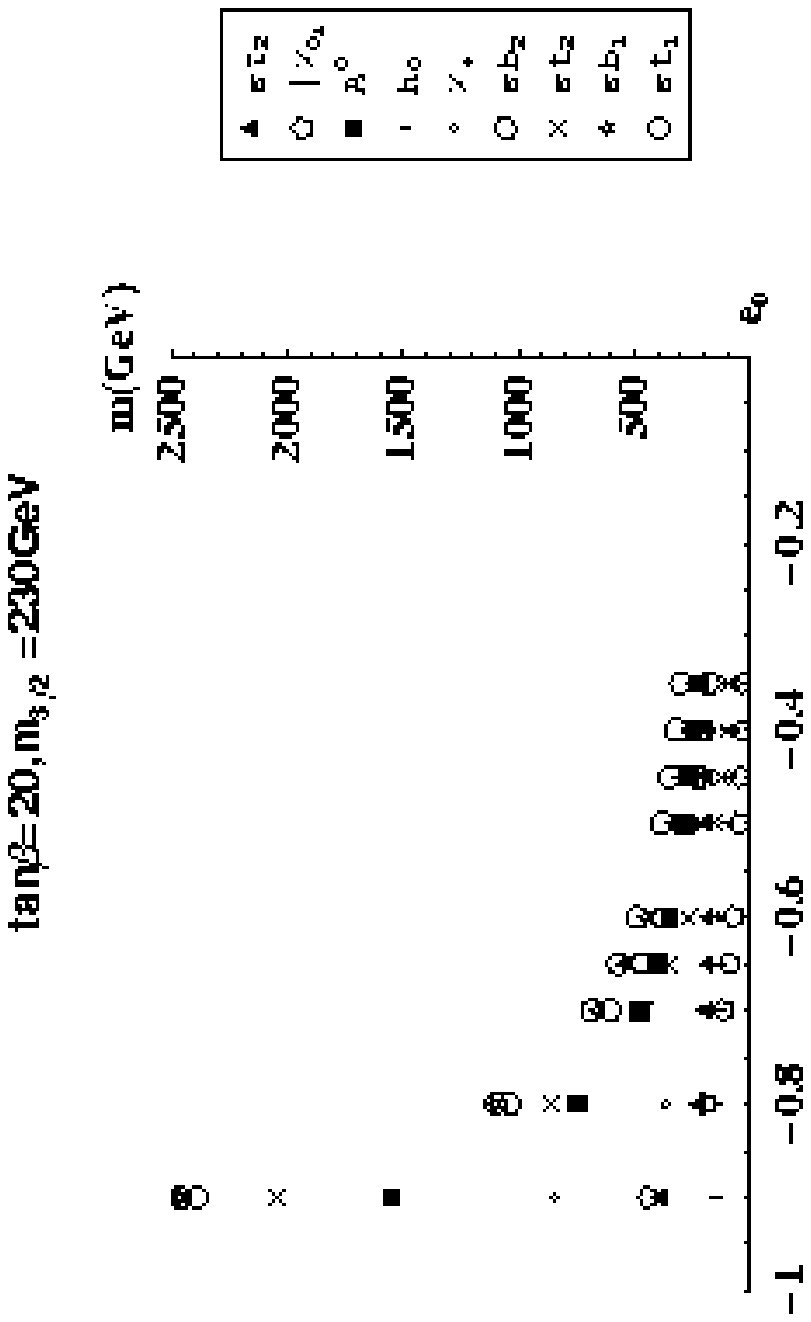}
\caption{Sparticle spectrum vs $e_O$ for $\tan\beta=20,m_{3/2}=230GeV$}
\end{figure}




\begin{thebibliography}{99}
\bibitem{HORWIT} P. Ho${\rm {\breve r}}$ava and E. Witten, Nucl. Phys. 
B460(1996)506; Nucl. Phys. B475(1996)94
\bibitem{witten} E. Witten, Nucl. Phys. B471(1996)135
\bibitem{ANTO} I. Antoniadis and M. Quir${\rm {\acute o}}$s, 
Phys.Lett.B392(1997)61;hep-th/9705037
\bibitem{NANO} T. Li, J.L. Lopez and D.V. Nanopoulos,hep-ph/9702237;
hep-ph/9704247; D.V. Nanopoulos, hep-th/9711080
\bibitem{LIS} T. Li, hep-th/9801123
\bibitem{NILLES} H.P. Nilles, M. Olechowski and M. Yamaguchi, hep-ph/9707143,
Phys. Lett.B 415(1997)24; H.P. Nilles, M. Olechowski and 
M. Yamaguchi, hep-th/9801030
\bibitem{DUDAS} E. Dudas and C. Grojean, hep-th/9704177;
E. Dudas,hep-th/9709043
\bibitem{STEVE} Z. Lalak and S. Thomas, hep-th/9707223
\bibitem{CHOI} K. Choi, H.B. Kim and C. Mu${\rm \tilde{n}}$oz, 
Phys. Rev. D57 (1998)7521, hep-th/9711158
\bibitem{LUKAS} A. Lukas, B.A. Ovrut and D. Waldram, hep-th/9711197;
A. Lukas, B.A. Ovrut and D. Waldram, hep-th/9710208
\bibitem{BURT}A. Lukas, B.A. Ovrut and D. Waldram, Phys.Rev. 
D59 (1999) 106005, hep-th/9808101 ;
A. Lukas, B.A. Ovrut and D. Waldram, hep-th/9901017;
JHEP 9904 (1999)009; R. Donagi, A. Lukas, B.A. Ovrut and D. Waldram, 
hep-th/9901009
\bibitem{BKL} D. Bailin, G.V. Kraniotis and A. Love, Phys.Lett.B432(1998)90;
hep-ph/9812283,  To appear in {\it Nuclear Physics B}
\bibitem{KINE}C.E. Huang, T. Liao, Q. Yan and T. Li, hep-ph/9810412  
\bibitem{CARLOS} S. Abel and C. Savoy, Phys.Lett. B444 (1998) 119-126,
hep-ph9809498 
\bibitem{CASA} J.A. Casas, A. Ibarra and C. Munoz,hep-ph/9810266
\bibitem{LI} T. Li, MADPH-99-1109, hep-ph/9903371
\bibitem{TATSUO} T. Kobayashi, J. Kubo and H. Shimabukuro, HIP-1999-14/TH,
hep-ph/9904201
\bibitem{MUNOZ} D.C. Cerdeno and C. Munoz, hep-ph/9904444
\bibitem{Tam:Rad} L. Ibanez, G.G. Ross: Phys. Lett. B110 (1982) 215;
L. Alvarez-Gaume, J. Polchinsky and M. Wise: Nucl. Phys. B221 (1983)
495; J. Ellis, J.S. Hagelin, D.V. Nanopoulos and K. Tamvakis: Phys.
Lett. B125 (1983) 275; M. Claudson, L. Hall and I. Hinchliffe: Nucl.
Phys. B228 (1983) 501; C. Kounnas, A.B. Lahanas, D.V. Nanopoulos and M.
Quiros: Nucl. Phys. B236 (1984) 438; L.E. Ib$\acute{a}$$\tilde{n}$ez 
and C. Lopez: Nucl.
Phys. B233 (1984) 511; A. Bouquet, J. Kaplan, C.A. Savoy: Nucl. Phys.
B262 (1985) 299
\bibitem{IBA:Spain} A. Brignole, L.E. Ib$\rm {\acute{a}}\rm{\tilde{n}}$ez and 
C. Mu$\rm {\tilde{n}}$oz, Nucl. Phys.B422(1994)125
\bibitem{KOREA} K. Choi, H.B. Kim and H. Kim, Mod. Phys. Lett. A14(1999)125
\bibitem{NATH} P. Nath and R. Arnowitt, Phys. LettB289(1992)368
\bibitem{EUC} Review of Particle Physics, The European Physical Journal C,
Vol.3,(1-4)(1998)
\bibitem{MAS} D. Bailin, G.V. Kraniotis and A. Love in preparation.

\end{thebibliography}
\end{document}